\documentclass[preprint %
 aip,
 amsmath,amssymb,
 reprint,%
]{revtex4-1}

\usepackage{graphicx}% Include figure files
\usepackage{dcolumn}% Align table columns on decimal point
\usepackage{bm}% bold math
\usepackage[utf8]{inputenc}
\usepackage[T1]{fontenc}
\usepackage{mathptmx}
\usepackage{etoolbox}

\makeatletter
\def\@email#1#2{%
 \endgroup
 \patchcmd{\titleblock@produce}
  {\frontmatter@RRAPformat}
  {\frontmatter@RRAPformat{\produce@RRAP{*#1\href{mailto:#2}{#2}}}\frontmatter@RRAPformat}
  {}{}
}%
\makeatother
\begin{document}

\preprint{AIP/123-QED}

\title[]{Manipulating crack formation in air-dried clay suspensions with tunable elasticity}
% Force line breaks with \\
\author{Vaibhav Raj Singh Parmar}
 %\altaffiliation[Also at ]{Physics Department, XYZ University.}%Lines break automatically or can be forced with \\
\author{Ranjini Bandyopadhyay}%
 \email{ranjini@rri.res.in}
\affiliation{ 
Soft Condensed Matter Group, Raman Research Institute, C. V. Raman Avenue, Sadashivanagar, Bangalore, 560 080, India. %\\This line break forced with \textbackslash\textbackslash
}%

\date{\today}% It is always \today, today,
             %  but any date may be explicitly specified

\begin{abstract}
Clay, the major ingredient of natural soils, is often used as a rheological modifier while formulating paints and coatings. When subjected to desiccation, colloidal clay suspensions and clayey soils crack due to the accumulation of drying-induced stresses. Even when desiccation is suppressed, aqueous clay suspensions exhibit physical aging, with their elastic and viscous moduli increasing over time as their microscopic structures evolve due to time-dependent inter-particle screened electrostatic interactions. The rate at which aging progresses is estimated from the rate of evolution of the mechanical moduli and can be controlled by changing clay concentration or by incorporating additives. Since physical aging and evaporation should both contribute to the consolidation of drying clay suspensions, we attempt to manipulate the desiccation process \textit{via} alterations of clay and additive concentrations. For a desiccating sample with an accelerated rate of aging, we observe faster consolidation into a semi-solid state and earlier onset of cracks. We estimate the crack onset time, $t_c$, in direct visualization experiments and the elasticity of the drying sample layer, $E$, using microindentation in an atomic force microscope. We demonstrate that $t_c \propto \sqrt{\frac{G_c}{E}}$, where $G_c$, the fracture energy, is estimated by fitting our experimental data to a linear poroelastic model that incorporates the Griffith's criterion for crack formation. Our work demonstrates that early crack onset is associated with lower sample ductility. The correlation between crack onset in a sample and its mechanical properties as uncovered here is potentially useful in preparing crack-resistant coatings and diverse clay structures. 
\end{abstract}
\maketitle

\section{\label{sec:1}Introduction}
Crack formation is a complex phenomenon that is observed over a range of length scales in diverse systems such as monolayers of microbeads, old paintings, craquelure in pottery glaze, industrial coatings, colloidal crystals, crocodile head scales, dried lakes and columnar jointing in cooling lava~\cite{doi:10.1098/rsta.2012.0353,Routh_2013,PhysRevE.59.1408,doi:10.1021/cm061931y,10.1130/0016,10.1063/5.0173925}. Evaporating films, coatings and surface layers constituted by aqueous colloidal suspensions tend to shrink and crack~\cite{doi:10.1098/rsta.2012.0353,Routh_2013,KITSUNEZAKI2011311,10.1063/5.0153682}. Interestingly, differential shrinkage induced by non-uniform cooling also leads to cracking~\cite{doi:10.1098/rsta.2012.0353}. As water evaporates from a colloidal sample, the air-water interface at the free evaporation surface develops a meniscus, which enhances the internal pore pressure. Adhesion to a frictional substrate imposes constraints on the shrinkage of a desiccating sample, results in the build-up of tensile stresses and can eventually lead to the formation of cracks. Whether a soft material deforms or fractures due to the buildup of drying-induced stresses is determined by its ductility and rigidity~\cite{D3SM01740K}. 
Once a crack is nucleated, its growth kinetics is described by the Griffith's criterion~\cite{griffiths,10.1002/9783527671922}, with the released strain energy exceeding the energy required to create two new fracture surfaces. In brittle solids, the formation of new surfaces requires breaking the bonds between the constituents. Colloidal systems are typically ductile, with the formation of new surfaces requiring comparatively more energy due to the presence of dissipative processes such as the rearrangement of particles near the crack tip~\cite{PhysRevLett.110.024301}. Drying-induced cracks form successively, with a previously formed crack determining the stress field governing the formation of a new crack~\cite{PhysRevE.71.046214}. In such hierarchical crack patterns, a new crack meets an existing one at 90$^\circ$. In contrast, for very thin drying layers~\cite{Groisman_1994} and samples undergoing repeated drying and wetting cycles ~\cite{B922206E}, crack joint angles of 120$^\circ$ have been reported.

Though undesirable in many applications, understanding drying-induced cracks is useful in the investigation of ancient geophysical processes~\cite{doi:10.1073/pnas.0805132106}, for identifying health problems by drying drops of blood~\cite{10.1115/1.4006033} and as a forensic tool for monitoring drying blood pools in crime scenes~\cite{LAAN2016104}. Experiments and theoretical analyses have demonstrated that colloidal suspension layers are a good model system for understanding crack formation in old paintings \cite{C7SM00985B}. The morphologies and growth kinetics of desiccation cracks in drying colloidal samples depend on the evaporation rate, initial volume fraction, rheology and thickness of the sample, shapes, sizes and surface charges of the constituent colloidal particles, substrate-sample adherence, substrate wettability and elasticity, applied fields and shape of the confining boundary~\cite{TANG2010261,Giorgiutti-Dauphin2014,PhysRevE.59.3737,Nag_2010,Groisman_1994,SANTANACHCARRERAS2007160,Thiery2016,PhysRevMaterials.2.085602,D1SM00820J,PhysRevE.103.032602, doi:10.1021/acs.langmuir.2c00197}. 
Desiccation cracks in soils are particularly interesting due to their geophysical relevance~\cite{10.1002/9783527671922,TANG2021103586,doi:10.1021/ie071375x}. Clay, a major ingredient of natural soil, exhibits very interesting rheology when mixed in water, with the initially liquid-like clay suspension losing its ability to flow and evolving gradually into a viscoelastic solid \textit{via} a physical aging mechanism~\cite{claybook,PhysRevLett.93.228302,PhysRevE.64.021510,Knaebel_2000}. While cracking because of solvent loss has been studied in great detail, the effect of physical aging of the sample on its drying and cracking behaviours has received less attention.

We report systematic experiments on desiccation-induced crack formation in thin aqueous layers of the synthetic clay Laponite$^\text{\textregistered}$. In powder form, monodisperse disk-shaped Laponite clay nanoparticles of 25 nm diameter and 1 nm thickness are arranged in one-dimensional stacks or tactoids.
When this powder is mixed with water, the soluble Na$ ^+$ ions that reside in the interlayer gallery spaces migrate towards the bulk aqueous medium due to osmotic pressure gradients~\cite{claybook}. To maintain overall charge neutrality, the faces of the clay particles acquire negative charges. Screened electrostatic double layer repulsions between the faces of like-charged clay particles within the tactoids result in tactoid swelling and exfoliation~\cite{doi:10.1021/acs.langmuir.8b01830,C0SM00590H}, as illustrated in supplementary material Fig.~S1(a).
 Diffuse electric double layers of Na$^+$ ions that form around the negatively charged clay particles continue to evolve with time as tactoid exfoliation progresses gradually~\cite{doi:10.1021/acs.langmuir.5b00291,ALI201585}. If the medium pH $<$ 11, the clay particle edges acquire slight positive charges due to the hydration of magnesia groups ~\cite{TAWARI200154,C2SM25731A}.
Individual disk-shaped clay particles in aqueous suspensions therefore have negative faces and positive rims, and spontaneously self-assemble into overlapping coin and house of cards configurations, as shown in supplementary material Figs.~S1(b,c). The gradual formation and restructuring of system-spanning fragile clay gel networks result in a continuous evolution in suspension viscoelasticity~\cite{doi:10.1021/acs.langmuir.8b01830,C0SM00590H,C2SM25731A,D2SM01457B}. Systematic studies of the evolution of the mechanical moduli of aging clay suspensions~\cite{PhysRevLett.93.228302,PhysRevE.64.021510,Knaebel_2000} due to the incorporation of additives~\cite{doi:10.1021/acs.langmuir.5b00291,THRITHAMARARANGANATHAN2017304,D1SM00987G}, application of electric fields~\cite{C8SM00533H,C5TB00506J,doi:10.1021/la702989u} and by tuning the solvent temperature and pH~\cite{TAWARI200154,D1SM00987G} are available in the literature.

Using direct visualization experiments and digital image correlation analyses, we recorded the time, $t_c$, and the location at which each desiccating clay sample first cracked. In the presence of adequate sample-boundary adhesion, we observed that the first crack always appeared near the vertical boundary of the sample cell, where the differential strain was highest. 
Furthermore, we demonstrated that adding common salt to the aqueous suspension medium accelerated sample consolidation and promoted crack formation, while adding tetrasodium pyrophosphate slowed down sample consolidation and inhibited crack onset.  %Furthermore, we demonstrated that crack formation could be promoted (or inhibited) by accelerating (or delaying) the sample consolidation by adding common salt (or tetrasodium pyrophosphate) to the aqueous suspension medium. 
We measured the elasticity values, $E$, of partially dried clay samples with and without common salt by performing microindentation experiments in an atomic force microscope. 
%We explained the observed preferential propagation of the cracks tangential to the vertical boundary by analyzing the radial stress using the theory of linear poroelasticity. 
Incorporating the Griffith's criterion for crack propagation into the poroelastic model, we predicted a simple scaling relationship, $t_c \propto \sqrt{\frac{G_c}{E}}$, between the crack onset time, $t_c$, sample elasticity, $E$, and fracture energy, $G_c$, of a drying clay sample at $t = t_c$. Our experimental data, acquired from desiccating clay suspensions with different initial clay and salt contents, satisfied this theoretical scaling relation and allowed us to estimate sample ductility and the likelihood of fracture onset in the samples. We concluded that samples with higher common salt and clay content exhibited relatively lower ductility and, therefore, reduced ability to sustain large deformations. Interestingly, the observed preferential propagation of the cracks tangential to the vertical boundary could also by rationalized using the theory of linear poroelasticity by analyzing the radial stresses on the surface of the drying clay sample.

\begin{figure*}[t]
    \centering
    \includegraphics[width=0.99\textwidth]{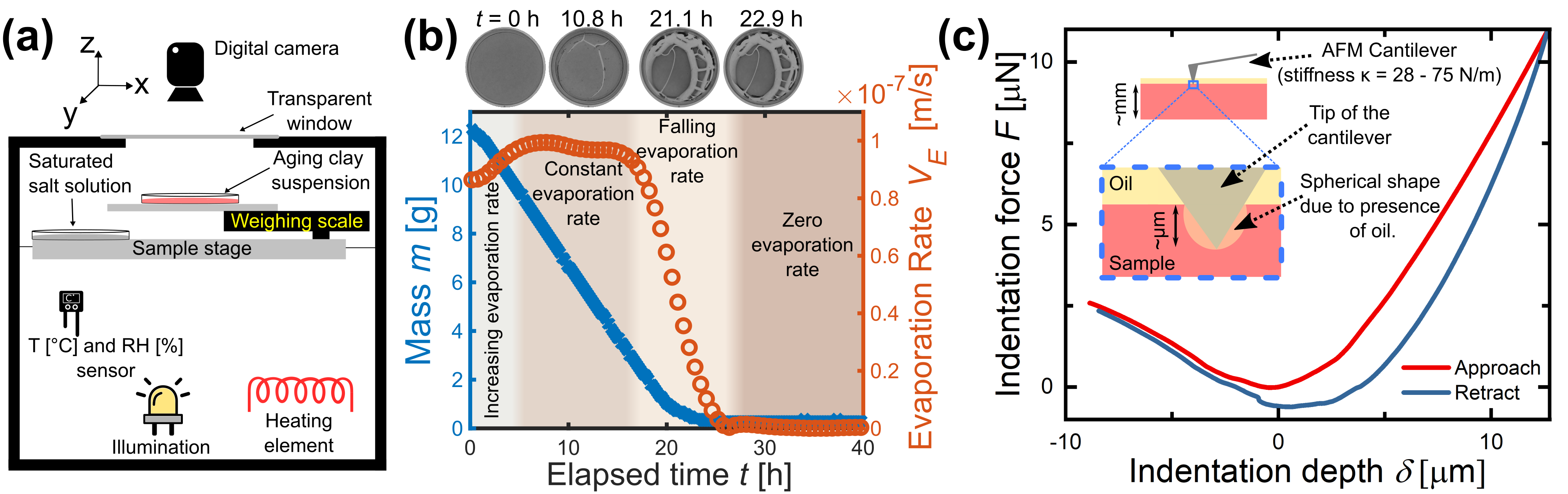}
    \caption{\textbf{(a)} Schematic diagram of the experimental chamber for imaging the onset and growth of desiccation cracks. \textbf{(b)} The top panel displays grayscale images, acquired at 4 different times, showing the onset and propagation of cracks in a drying clay suspension of clay concentration $C_L$ = 2.5\% w/v and NaCl concentration $C_A$ = 4 mM. Mass $m$ (left y-axis, blue) of the drying sample and the corresponding evaporation rate $V_E$ (right y-axis, orange) as a function of elapsed time $t$. \textbf{(c)} A representative force-indentation depth curve for a partially dried Laponite suspension of $C_L$ = 2.5\% w/v and $C_A$ = 2 mM NaCl, obtained from microindentation measurements in an atomic force microscope.}
    \label{fig1}
\end{figure*}
\section{\label{sec:2}Materials and methods}
\subsection{\label{sec:2.1}Sample preparation and characterization of mechanical properties}
Hygroscopic Laponite$^\text{\textregistered}$ XLG powder was procured from BYK Additives Inc. and baked at 120$^\circ$C for $18-24$ hours to remove moisture. Aqueous solutions were prepared by adding sodium chloride, NaCl (common salt, LABORT Fine Chem Pvt. Ltd), or tetrasodium pyrophosphate, TSPP (E. Merck (India) Ltd), to deionized and distilled water (Millipore Corp., resistivity 18.2 M$\Omega$-cm). Clay samples with and without additives were prepared by slowly adding predetermined amounts of dried Laponite$^\text{\textregistered}$ powder to 50 ml Milli-Q water and to aqueous solutions prepared with additives (NaCl or TSPP) under vigorous stirring conditions.
Homogeneous suspensions with clay concentrations 2.0, 2.5, 3.0 and 3.5\% w/v were prepared using this method. After 40 minutes of vigorous mixing of the transparent clay suspension, 200 $\mu$l of 2.09 mM Rhodamine B dye was added to enhance contrast in imaging studies. Finally, the dyed clay suspensions were stirred for an additional 5 minutes. The evolution of mechanical properties of aging clay suspensions was studied at 25$^\circ$C in a double gap geometry (DG26.7) by performing oscillatory rheological measurements at 0.1\% strain in an Anton Paar MCR702 rheometer.

\subsection{\label{sec:2.2}The desiccation experiment} 
12 ml of the freshly prepared clay suspension was poured into a transparent polymethylpentene (TPX) Petri dish with an inner diameter of 4.65 cm. The initial clay layer thickness was kept constant at 7 mm in all the desiccation experiments reported here. The Petri dish was placed immediately in a temperature and humidity-controlled chamber, a schematic illustration of which is presented in Fig.~\ref{fig1}(a). The temperature in the chamber was maintained at 35, 40, 45 and 50 $^\circ$C in separate experimental runs using Arduino-based relay switching of a heating element. Humidity was maintained at 27 $\pm$ 4 \% by placing a saturated salt solution of magnesium chloride in the chamber~\cite{Greenspan1977-sl}. The variations in temperature and relative humidity inside the chamber during the experiment, shown in supplementary material Fig.~S2, were minimal. Images of the top surface of the drying suspension layer were captured using a Nikon D5600 camera, with consecutive images taken at 3-minute intervals. Depending on the chamber temperature, the total drying duration of all the samples studied here varied between 22 and 46 hours.
\subsubsection{\label{sec:2.2.1}Imaging crack onset and propagation}
 A representative video of the temporal evolution of a desiccating sample with clay concentration of 3.0\% w/v and NaCl concentration of 4mM is presented in Supplementary Video 1. The acquired images were converted to grayscale format and analyzed using the MATLAB@2021 image processing toolbox. These images were binarized after setting a threshold. We recorded the time of crack onset, $t_c$, by determining the instant at which the substrate was first exposed due to the appearance of the first crack. The procedure for determining $t_c$ is schematically illustrated in supplementary material Fig.~S3. The Petri dish containing the drying sample was placed on an Arduino-controlled weighing scale with a transparent acrylic measuring plate and the mass of the drying sample was monitored at intervals of 10 minutes. The blue symbols in Fig.~\ref{fig1}(b) represent the observed monotonic decrease in the mass of the aqueous sample till the completion of evaporation. The evaporation rate, $V_E$, was determined from the change in sample mass due to loss of solvent, using $V_E = \frac{dm/dt}{\pi \rho_w R^2}$~\cite{https://doi.org/10.1002/2013WR014466}. Here, $dm/dt$ is the sample mass loss per unit time, $\rho_w$ = 1 g/cm$^3$ is the density of water and $\pi R^2$ is the area occupied by the evaporating clay surface in the Petri dish of radius $R$ = 2.325 cm. The temporal variation of the evaporation rate, $V_E$, of the desiccating sample is also shown by orange symbols in Fig.~\ref{fig1}(b). The sample, prepared at room temperature, was initially equilibrated to 50$^\circ$C, after which distinct evaporation zones were noted. After maintaining a constant value for a few hours, $V_E$ decreased rapidly to zero as the sample dried up. Distinct evaporation regimes were also reported for drying nanopore gels~\cite{Thiery2016} and wet granular media~\cite{Coussot2000}. As in earlier observations~\cite{TANG2021103586,TRAN2019142}, the first crack always emerged during the constant evaporation rate regime in all our desiccating samples. {This implies that all the clay suspensions were fully saturated with water ~\cite{SCHERER198877} at the time of crack nucleation.}

 \subsubsection{\label{sec:2.2.2}Strain field analysis using digital image correlation}
 A random texture was created by sprinkling fine black pepper powder over the surface of the drying clay samples. Texture deformation due to desiccation-induced strains was imaged and analyzed using an open-source 2D digital image correlation (DIC) Matlab software `Ncorr'~\cite{Blaber2015,DICalgo}. Ncorr determines one-to-one correspondences between a reference and subsequent images using linear combinations of translation, shear and stretch transformations to estimate the deformation fields in the drying sample. This protocol was used to evaluate the diagonal components of the two-dimensional strain tensor and estimate the extents of elongative and compressive stresses on the sample surface. DIC analyses were performed using highly elastic Laponite$^\text{\textregistered}$ suspensions of particle concentration 3.0\% w/v and NaCl concentration 4 mM to ensure only deformation and no flow.

 \subsection{\label{sec:2.3} Determining sample elasticity by microindentation using atomic force microscopy}
 The elasticity of each clay sample was determined at the time of onset of the first crack, $t = t_c$, and at $t = 14$ h using a microindentation technique~\cite{Gavara2016} in an atomic force microscope (Asylum Research, Oxford Instruments). Since clay samples contain water, microindentation can accurately measure the force with changing indentation depth only when the capillary attraction between the cantilever and sample is minimal. As illustrated in the inset of Fig.~\ref{fig1}(c), a low capillary attraction condition was achieved by applying a thin layer of silicon oil (Sigma-Aldrich) of viscosity $\eta = 5$~cSt on the surface of the partially dried sample. A cantilever of stiffness 28 - 75 N/m (Tap190-G, Budget Sensors), capable of withstanding the capillary attraction with the sample while still staying below the AFM trigger point voltage, was used to perform microindentation measurements. This additional step in the measurement protocol enabled us to accurately determine the elasticity of aqueous clay suspensions and can be extended to other water-containing viscoelastic samples. Details of the methodology adopted for extracting sample elasticity from force-indentation depth curves are presented in supplementary material Section~ST1 and Fig.~S4. 
Indentation experiments could not be performed for clay samples with added TSPP due to immeasurably low elasticity values. 

   Even though the tip of the cantilever used in these experiments is pyramid-shaped, the oil on the cantilever is expected to assume a spherical shape due to surface tension, as illustrated in the inset of Fig.~\ref{fig1}(c). A representative force-indentation depth curve at elapsed time $t   \approx t_c$ = 10.25 h for a partially dried clay suspension of concentration 2.5\% w/v with 2 mM NaCl is shown in Fig.~\ref{fig1}(c). The force-indentation depth curve measured for each sample was fitted with the Hertz contact model for a spherical tip during the cantilever approach cycle. The measured force, $F(\delta)$, at indentation depth $\delta$, given by $F(\delta) = \frac{4 E \sqrt{R_c} \delta^{3/2}}{3(1 - \nu^2)}$, was used to extract sample elasticity, $E$. Here, $R_c = 10$ nm is the cantilever-tip radius. Since the Poisson's ratio $\nu$ for colloidal gels and saturated clay ranges between 0.4 and 0.5~\cite{TAKIGAWA19961,C1SM06484C,ARACHCHIGE2021228773}, we have assumed $\nu$ = 0.5 while computing $E$. We have verified that selecting any value of $\nu$ between 0.4 and 0.5 does not qualitatively change our subsequent results. Force-indentation depth data for a standard polymer sample PDMS, acquired using microindentation and fitted to the Hertz model is displayed in supplementary material Fig.~S5. This analysis yielded an elasticity value of 1.62 MPa, in agreement with the available literature.

   \section{\label{sec:3}Results and discussions}
   \begin{figure}[ht]
    \centering
    \includegraphics[width=0.49\textwidth]{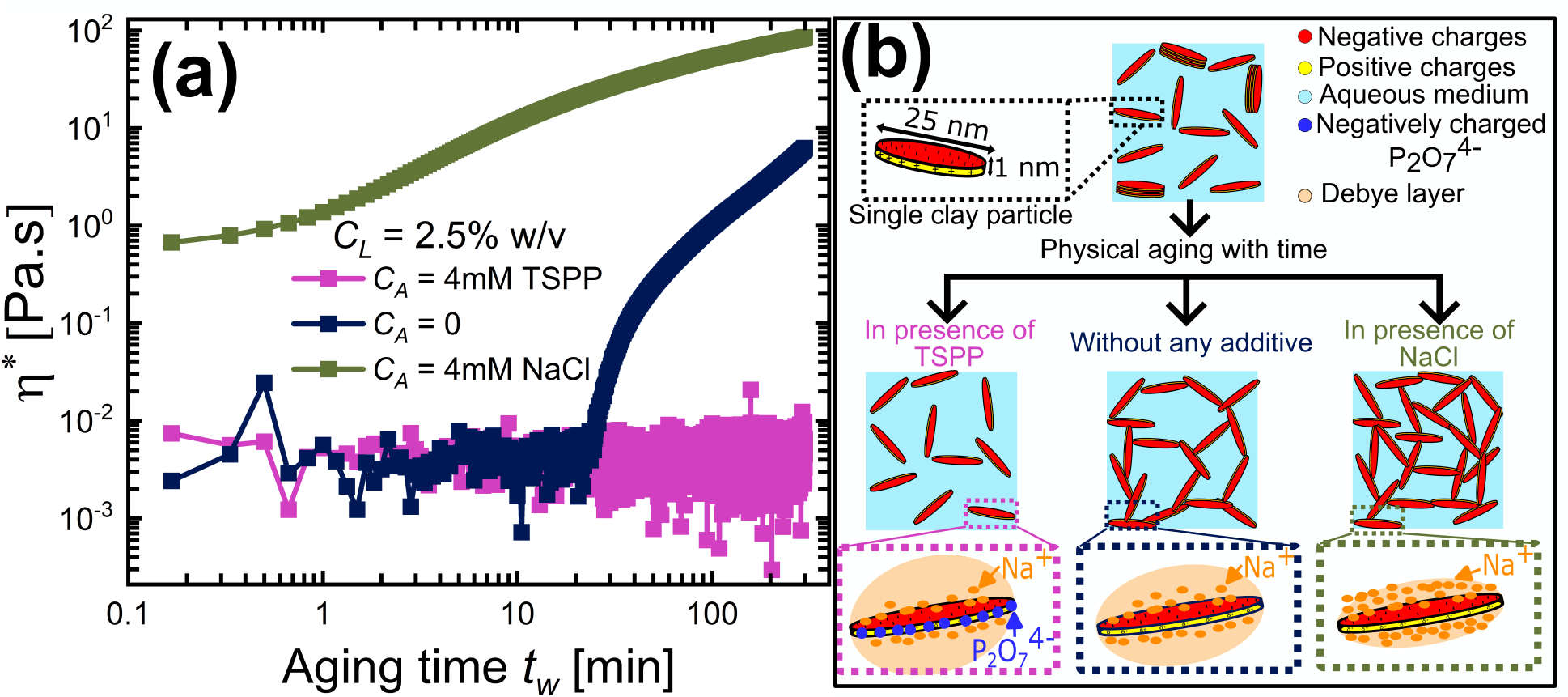}
    \caption{\textbf{(a)} Temporal evolution of complex viscosity, $\eta^* = \sqrt{G^{\prime^2} + G^{\prime\prime^2}}/\omega$, in aqueous Laponite$^\text{\textregistered}$ suspensions of $C_L =$ 2.5\% w/v with no additive ($C_A =$ 0, navy blue), $C_A =$ 4mM NaCl (olive green) and $C_A =$ 4 mM TSPP (fuscia) at 25$^\circ$C. $G^{\prime}$ and $G^{\prime\prime}$ are, respectively, the elastic and viscous moduli of the suspension and were measured in rheology experiments at a strain amplitude $\gamma$ = 0.1\% and angular frequency $\omega$ = 6 rad/sec. \textbf{(b)} Schematic illustration of the self-assembly of clay particles in suspensions with and without additives.}
    \label{fig2}
\end{figure}
\subsection{Physical aging of aqueous Laponite$^\text{\textregistered}$ clay suspensions}
Figure~\ref{fig2}(a) shows the distinct time-evolution behaviours of the complex viscosity $\eta^*=\sqrt{G^{\prime^2} + G^{\prime\prime^2}}/\omega$ for clay suspensions of concentration 2.5\% w/v with and without additives. 
The gradual formation and restructuring of fragile clay gel networks leads to a physical aging process, which is observed as a continuous increase in both the elastic and viscous moduli of the clay suspension~\cite{doi:10.1021/acs.langmuir.8b01830,C2SM25731A,C0SM00590H} in the absence of additives and when NaCl is added to the suspension medium. Enhancements in the elastic and viscous moduli contribute to an overall increase in the complex viscosity, $\eta^*$. When common salt (NaCl) was added to the clay suspension, the evolution of complex viscosity $\eta^*$ was the most rapid, indicating accelerated gel network formation~\cite{doi:10.1021/acs.langmuir.5b00291,D1SM00987G}. The Debye layer surrounding each clay particle shrinks in the presence of the Na{$^+$} and Cl{$^-$} ions, thereby reducing interparticle repulsion and enhancing the formation of self-assembled gel networks~\cite{C2SM25731A}. When tetrasodium pyrophosphate (TSPP) was introduced to the medium, the attachment of {P$_2$O$_7$}$^{4-}$ groups to clay particle rims enhanced interparticle repulsions~\cite{PhysRevE.66.021401}. In Fig.~\ref{fig2}(a), the low and approximately constant values of $\eta^*$ in the presence of TSPP arise from the absence of system-spanning particulate structures in these repulsive suspensions. These results are in agreement with previous studies~\cite{doi:10.1021/acs.langmuir.5b00291,PhysRevE.66.021401}. The physical aging mechanisms by which clay suspension microstructures evolve in the presence of NaCl and TSPP have been illustrated schematically in Fig.~\ref{fig2}(b). Clay particle self-assembly, sample aging dynamics and the rate of evolution of mechanical properties can therefore be tuned by incorporating suitable additives. Furthermore, the aging dynamics of clay suspensions, and therefore their mechanical moduli, can also be accelerated by increasing clay concentration~\cite{doi:10.1021/acs.langmuir.5b00291}. 

\begin{figure}[ht]
    \centering
    \includegraphics[width=0.49\textwidth]{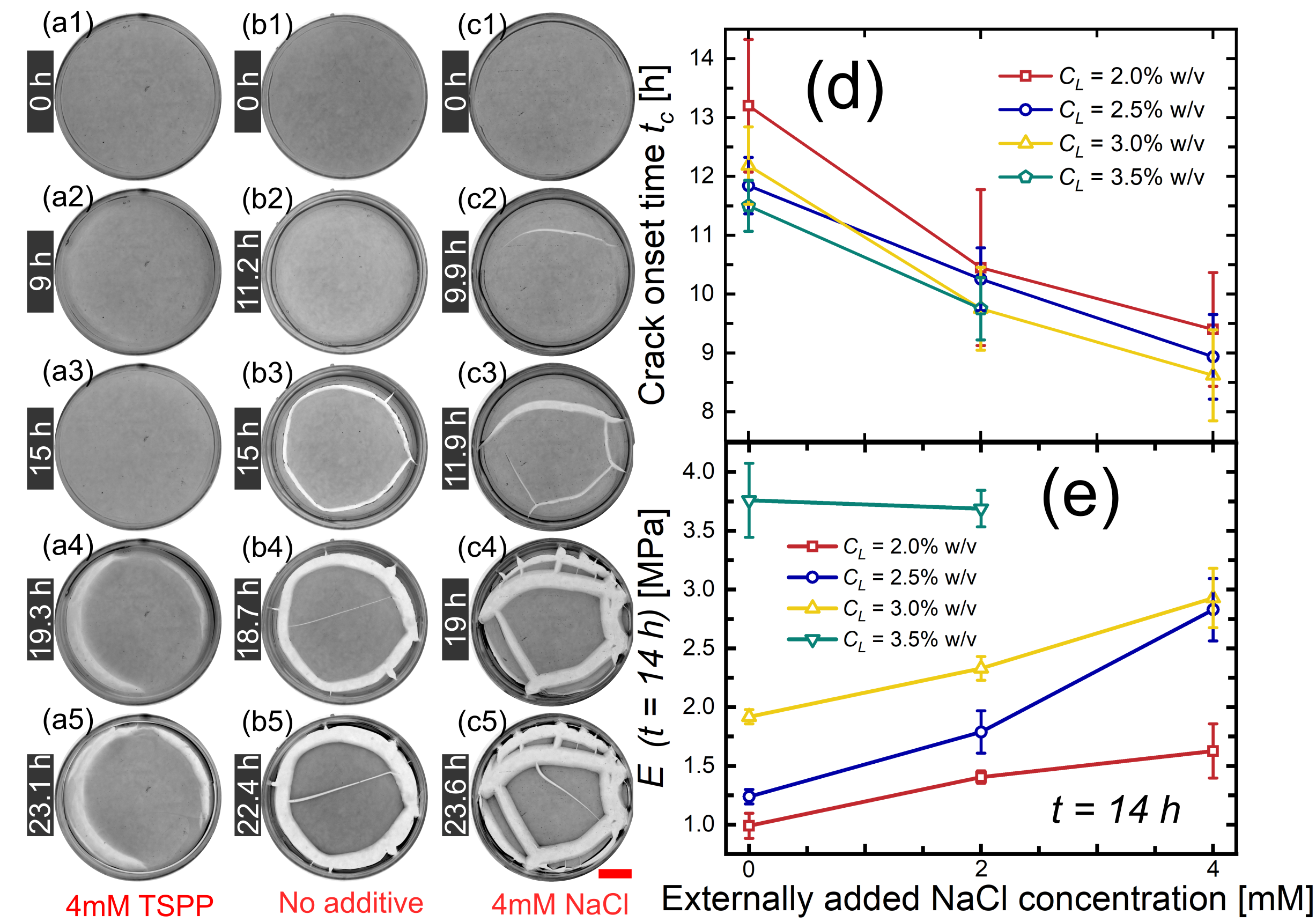}
    \caption{Time evolution of desiccation cracks in aging Laponite clay suspensions of $C_L$ = 2.5\% w/v as the solvent evaporated in the presence of \textbf{(a1-a5)} $C_A =$ 4 mM TSPP, \textbf{(b1-b5)} $C_A =$ 0 and \textbf{(c1-c5)} $C_A =$ 4 mM NaCl. The time elapsed since suspension loading is mentioned on the left of each subfigure. The scale bar at the bottom right of (c5) is 1 cm. \textbf{(d)} The crack onset time, $t_c$, defined as the time elapsed when the first crack forms and \textbf{(e)} elasticity of each sample at a constant elapsed time $t = 14$ h \textit{vs.} NaCl concentration for different clay concentrations, $C_L$. The error bars represent standard errors in estimating $t_c$ and $E$($t = 14$ h) respectively, and were calculated from 3-4 direct visualization and 4-8 microindentation experiments.
    }
    \label{fig3}
\end{figure}

   \subsection{Faster consolidation of clay suspensions leads to earlier onset of cracks}
In our first set of desiccation experiments, the evaporation rate of the sample was fixed at $V_E$ = 9.75 x 10$^{-8}$ m/s by maintaining constant temperature and relative humidity in the experimental chamber.
The aging dynamics of the sample, and therefore the rate of buildup of self-assembled suspension structures, was tuned by controlling clay and additive concentrations.
Figures~\ref{fig3}(a-c) display the temporal evolution of desiccation cracks in drying clay suspensions, prepared at a concentration of 2.5\% w/v, both with and without additives.
Addition of TSPP resulted in lateral shrinkage of the drying sample, which detached from the Petri dish boundary without the formation of desiccation cracks (Figs.~\ref{fig3}(a1-a5)). For the clay suspension without additive, cracks formed before the occurrence of lateral shrinkage (Figs.~\ref{fig3}(b1-b5)). In the presence of NaCl, the first crack appeared even earlier and cracking increased substantially (Figs.~\ref{fig3}(c1-c5)).  Crack onset time \textit{vs.} added NaCl concentration is plotted in Fig.~\ref{fig3}(d) for different clay concentrations. It is seen that cracks formed earlier with increase in salt and clay concentrations due to increase in suspension elasticity driven by accelerated aging. Although the first crack always formed in the constant evaporation rate regime, there is a large standard deviation in the crack onset times, presumably due to local strain field heterogeneities in the drying sample. %Clearly, solvent loss due to evaporation and buildup of suspension structures due to physical aging both contributed to the consolidation of the drying clay sample.% 
To verify the relative contributions of solvent loss and physical aging  %and solvent loss due to evaporation 
in the consolidation of the drying clay suspensions studied here, we measured the elasticity values characterizing partially dried clay samples in fully developed gel states at a constant elapsed time $t = 14$ h using microindentation in an atomic force microscope. It is seen from Fig.~\ref{fig3}(e) that the elastic moduli, $E$($t = 14$ h), of clay samples at a constant elapsed time is higher when the NaCl and clay contents are higher. Cracking is therefore more extensive in samples that are characterized by faster aging. Since the evaporation rates for all clay samples are identical, this observation indicates that the contribution of physical aging in driving sample consolidation is extremely significant. 

 The final crack patterns, imaged after the completion of the desiccation process, are shown in supplementary material Fig.~S6(a). The exposed substrate area quantifies the extent of drying and was estimated from the area of the bottom surface of the Petri dish that was visible after the completion of drying. As shown in supplementary material Figs.~S6(b,c), the exposed substrate area was greater for samples with higher NaCl content, indicating more extensive cracking and shrinking.
 Since the consolidation rate of clay suspensions can also be varied by changing the evaporation rate during drying, crack onset times for drying suspensions with 3\% w/v clay and 4 mM NaCl were recorded at different temperatures and therefore at different evaporation rates, $V_E$ (supplementary material Figs.~S7(a,b)). The data presented in supplementary material Fig.~S7(c) shows an inverse correlation between crack onset time and evaporation rate. This observation agrees with previous reports~\cite{Giorgiutti-Dauphin2014,SCHERER198877}.

\subsection{Strain field maps reveal that differential strains are largest near the vertical boundary of the Petri dish}
\begin{figure}[t]
\includegraphics[width=0.49\textwidth]{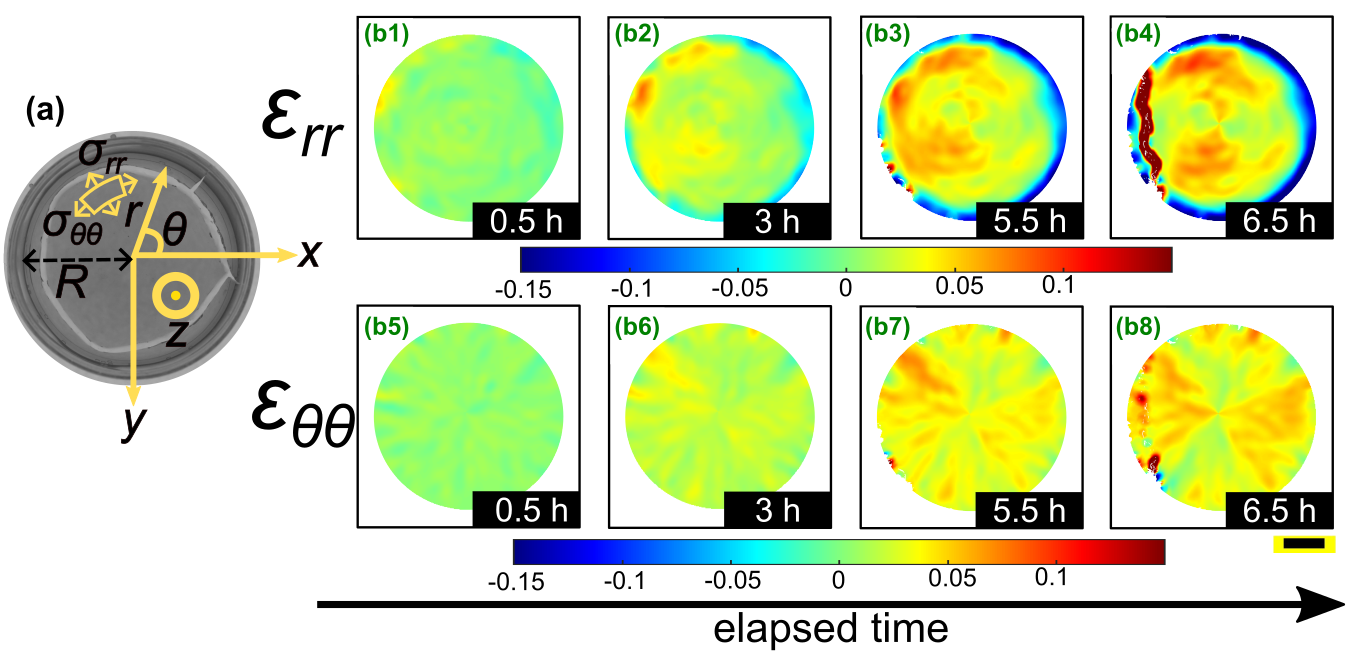}
\centering
    \caption{\textbf{(a)} Illustration of the {sample} geometry (top view) showing the cylindrical coordinate system used in this study. \textbf{(b1-b8)} Temporal evolution of strain fields along the $r$ and $\theta$ directions, $\epsilon_{rr}$ and $\epsilon_{\theta\theta}$, \textit{vs.} elapsed time for a drying Laponite$^\text{\textregistered}$ clay suspension of $C_L =$ 3.0\% w/v and with $C_A$ = 4 mM NaCl. The colour bars show the magnitudes of $\epsilon_{rr}$ and $\epsilon_{\theta\theta}$, with negative and positive strain values indicating compressive and elongative strains respectively. The time stamps at the bottom right of each sub-figure indicate the elapsed time. The scale bar at the bottom right of (b8) is 1 cm.}
    \label{fig:DIC}
\end{figure}

Figure~\ref{fig:DIC} displays the strain fields on the surface of a drying clay sample of concentration 3.0\% w/v with 4 mM added NaCl, analyzed from the acquired snapshots using digital image correlation. {Figure~\ref{fig:DIC}(a) describes the geometry of a clay layer in a circular Petri dish in cylindrical coordinates.}
 Figures~\ref{fig:DIC}(b1-b4) and ~\ref{fig:DIC}(b5-b8) show the evolutions of the diagonal components, $\epsilon_{rr}$ and $\epsilon_{\theta\theta}$, of the two-dimensional strain tensor till the formation of the first crack. Estimates of elongative (positive values) and compressive  (negative values) strains along the $r$ and $\theta$ directions, with the first image frame as reference, are represented by colour bars at the bottom of each row of sub-figures. Figs.~\ref{fig:DIC}(b1-b8) reveal that changes in $\epsilon_{rr}$ were far more significant than in $\epsilon_{\theta\theta}$. It is seen from Figs.~\ref{fig:DIC}(b1-b4) that while strong compressive strains developed near the vertical boundary as drying progressed, elongative strains were comparatively more uniformly distributed across the sample surface. In all the samples that we studied, the first crack formed near the vertical boundary of the Petri dish, where the buildup of adhesion-induced differential strains, $\Delta\epsilon_{rr}$, was maximum.

To further study the role of the boundary on crack formation, we reduced sample-boundary adhesion by coating the Petri dish with vacuum grease. Supplementary Video 2 shows the temporal evolution of drying clay suspensions for different sample-boundary adhesion conditions. The analyzed images, presented in supplementary material Fig.~S8 and discussed in supplementary material Section~ST2, show that reduction in sample-boundary adhesion promoted isotropic shrinkage and resulted in delayed cracking. For our subsequent studies on crack onset in drying clay suspension layers, we ensured maximum sample-boundary adhesion by loading the clay samples in uncoated Petri dishes.

\begin{figure*}[t]
\includegraphics[width=0.99\textwidth]{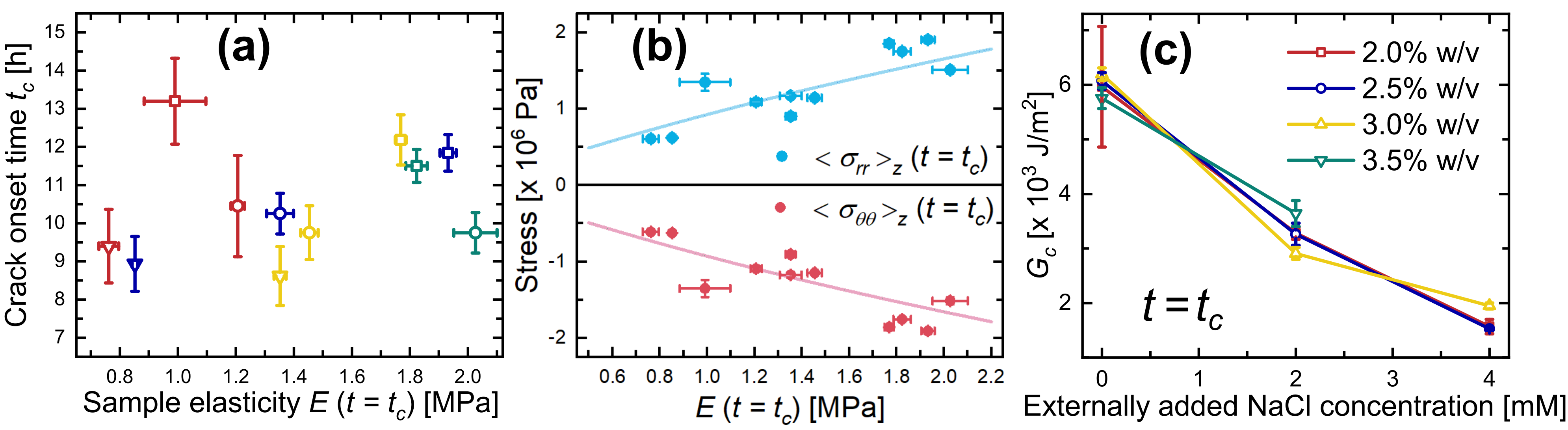}
\centering
    \caption{\textbf{(a)} Crack onset time $t_c$ as functions of sample elasticity, $E$, measured at $t = t_c$. Squares, circles and triangles correspond to NaCl concentrations $C_A =$ 0, 2 and 4 mM, respectively. Red, blue, yellow and green symbols correspond to clay concentrations $C_L =$  2.0, 2.5, 3.0 and 3.5\% w/v, respectively. \textbf{(b)} Height-averaged in-plane stresses, $<\sigma_{rr}>_z$ and $<\sigma_{\theta\theta}>_z$ \textit{vs.} $E$($t = t_c$) computed from experimental data. Solid lines are theoretical predictions of the in-plane stresses at the vertical boundary ($r = R$) of the Petri dish for an arbitrary sample with $h$ = $\Bar{h}$ = 3.7 mm and $t_c$ = $\Bar{t_c}$ = 10.5 h. Here ` $\Bar{}$ ' denotes the average values from our experimental data. \textbf{(c)} Fracture energy values, $G_c$, of the samples at $t=t_c$ as a function of externally added NaCl for different clay concentrations, calculated using experimental data and Eqns.~(3) and (4).}
    \label{fig5}
\end{figure*}
\subsection{Correlating crack onset time with rigidity and ductility of the drying clay layer}
Our experiments reveal that the first crack nucleates after the {drying} clay suspension layer reaches a semi-solid gel-like state. A network of individual clay particles constitutes this saturated gel, which is expected to be isotropic and could, in principle, be described by the linear poroelastic model. Drying-induced stresses are expected to accumulate {in} the sample due to a continuous increase in pore pressure drop.
To determine the tensile stress in the evaporating layer, we modeled the time-varying local liquid pore pressure in a cylindrical geometry represented by the coordinates ($r, \theta, z$), as shown in Fig.~\ref{fig:DIC}(a). We assumed that the consolidating clay suspension layer had an initially unfractured surface and the layer thickness decreased significantly due to desiccation. Following earlier reports on drying suspensions~\cite{10.1002/9783527671922,PhysRevE.103.032602,https://doi.org/10.1029/2010JF001842,D2SM00012A}, it is reasonable to expect that since the in-plane strains on the surface of the clay layer ($\epsilon_{rr}$ and $\epsilon_{\theta\theta}$) were constrained by the boundaries, the out-of-plane strain $\epsilon_{zz} >> \epsilon_{rr} + \epsilon_{\theta\theta}$. Furthermore, we assumed that the clay-air interface is traction-free ($\mathbf{\sigma.\hat{z}}$ = 0)~\cite{Chekchaki2013}
and, therefore, does not sustain stress, such that $\sigma_{zz} << \sigma_{rr} + \sigma_{\theta\theta}$. 

The local liquid pressure exerted on the pores of the clay-gel network under uniaxial strain and constant, minimal vertical stress should satisfy the linear poroelastic equation~\cite{10.1002/9783527671922,Giorgiutti-Dauphin2014}: 
\begin{equation}
    \frac{\partial p}{\partial t} = c \frac{\partial^2 p}{\partial {z^2}}
\end{equation}
In this equation, $c = \frac{\kappa E}{\eta}$ is the consolidation coefficient, where $\eta$ is the viscosity of water, $E$ is the elasticity and $\kappa$ is the permeability of the porous gel. As discussed in supplementary material Section~ST3, the permeability of our samples, estimated using the Carman–Kozeny relationship~\cite{Kruczek2015}, is $\approx 10^{-19} m^2$. Assuming $\eta = 10^{-3}$ Pa-s and $E \approx 10^6$ Pa, the consolidation coefficient $c$ $\approx 10^{-10}$ m$^2$/s, which indicates an extremely slow rate of water diffusion through the pores of the gel and inadequate compensation for solvent loss at the free evaporation surface. In addition, capillary pressure at the surface of the clay layer is expected to remain high due to meniscus formation. These factors, together with the adhesion of the sample with the Petri dish boundary, resulted in constrained lateral shrinkage and subsequent cracking.

The pressure across the thickness of the clay layer at the start of desiccation satisfies the initial condition $p(z,t = 0) = p_{atm}$, the atmospheric pressure. Since the first crack always nucleated during the constant evaporation rate regime, Darcy's law, given by $\frac{\partial p}{\partial z} = -\frac{\eta V_E}{\kappa}$ where $V_E$ is the constant evaporation rate, applies at the clay-air interface corresponding to $z = 0$. Solving Eqn.~(1) with these constraints gives the pore pressure as a function of $z$ and $t$~\cite{https://doi.org/10.1029/2010JF001842,D2SM00012A}:
\begin{equation}
    p(z,t) = p_{atm}-\frac{2V_E\eta}{\kappa} \left[ \left( \frac{ct}{\pi} \right)^{1/2}\exp\left(-\frac{z^2}{4ct}\right) + \frac{z}{2} \operatorname{erfc}\left(-\frac{z} {2\sqrt{ct}}\right) \right]
\end{equation}
 Here, $\operatorname{erfc}$ is the complementary error function. Using the approximations $\epsilon_{zz} >> \epsilon_{rr} + \epsilon_{\theta\theta}$,
$\sigma_{zz} << \sigma_{rr} + \sigma_{\theta\theta}$ and { the boundary condition $\sigma_{rr}|_{r=R} = p_{atm} - p(z,t)$}, the in-plane stress components, $\sigma_{rr}$ and $\sigma_{\theta\theta}$, in the clay suspension layer were analytically determined from the linear poroelastic constitutive relation. Identical strain and stress conditions were used earlier in modelling the drying of colloidal suspensions on rigid glass substrates and in drying drops of dense bacterial suspensions~\cite{PhysRevE.103.032602,D2SM00012A}. 
{Since the first crack always nucleated at $r = R$ in our experiments, the thickness-averaged in-plane stresses, calculated following a previous report~\cite{D2SM00012A} at $r=R$ for Poisson's ratio $\nu =$ 0.5, can be written as:}
\begin{equation}
\begin{split}
<\sigma_{rr}(t)|_{r = R}>_z = <p_{atm} - p(z,t)>_z = \frac{V_EEt}{h} \Biggl[1 + \frac{h\exp\left(-\frac{h^{2}}{4ct}\right)}{\sqrt{\pi ct}} \\
- \frac{h^2 + 2ct}{2ct} \operatorname{erfc}\left( \frac{h}{2}\sqrt{\frac{1}{ct}}\right)\Biggr] 
\end{split}
\end{equation}
%\begin{equation}
%\sigma_{\theta\theta} = <p_{atm} - p(z,t)>_z \frac{r^2(2\nu - 1) + R^2 \nu}%{r^2 (\nu - 1)}
%\end{equation}
and $<\sigma_{\theta\theta}|_{r = R}>_z$ = -$<\sigma_{rr}|_{r = R}>_z$. Here, $<p_{atm} - p(z,t)>_z = \int_{-h}^0 [p_{atm} - p(z,t)] dz$ is the average liquid pore pressure and $h$ is the thickness of the sample (layer height) at elapsed time $t$. %Given the angular stress $\sigma_{\theta\theta}$ = - $\sigma_{rr}$.
%\begin{equation} 
%\begin{split}
%<p_{atm} - p(z,t)>_z = \frac{V_EEt}{h} \Biggl[1 + \frac{h\exp\left(-\frac{h^{2}}{4ct}\right)}{\sqrt{\pi ct}} \\
%- \frac{h^2 + 2ct}{2ct} \operatorname{erfc}\left( \frac{h}%{2}\sqrt{\frac{1}{ct}}\right)\Biggr]
%\end{split} 
%\end{equation}

 To estimate the in-plane stresses {for clay samples at the time of crack onset}, we used $V_E$ = 9.75 x 10$^{-8}$ m/s, $\eta = 10^{-3}$ Pa-s, $\kappa \approx 10^{-19}$ m$^2$, while clay layer elasticity $E(t = t_c)$ was estimated using microindentation experiments in an AFM, and $h$($t = t_c$) was determined from the mass change of the sample due to solvent loss, as described in supplementary material Section~ST4. Fig.~\ref{fig5}(a) shows the absence of a clear correlation between $t_c$ and $E$($t = t_c$) for clay samples with different clay and NaCl concentrations. % for each data-point is mentioned in the brackets as the first and second element respectively. 
Fig.~\ref{fig5}(b) shows the calculated {height-averaged} in-plane stresses as a function of sample elasticity at $t = t_c$ for all the samples. The tensile nature of the radial stress ({positive} $<\sigma_{rr}|_{r = R}>_z$) and compressive nature of the angular stress ({negative} $<\sigma_{\theta\theta}|_{r = R}>_z$) resulted in the propagation of the first crack in a direction perpendicular to the tensile stress, and therefore, tangential to the vertical boundary. 

\begin{figure}[ht]
\includegraphics[width=0.49\textwidth]{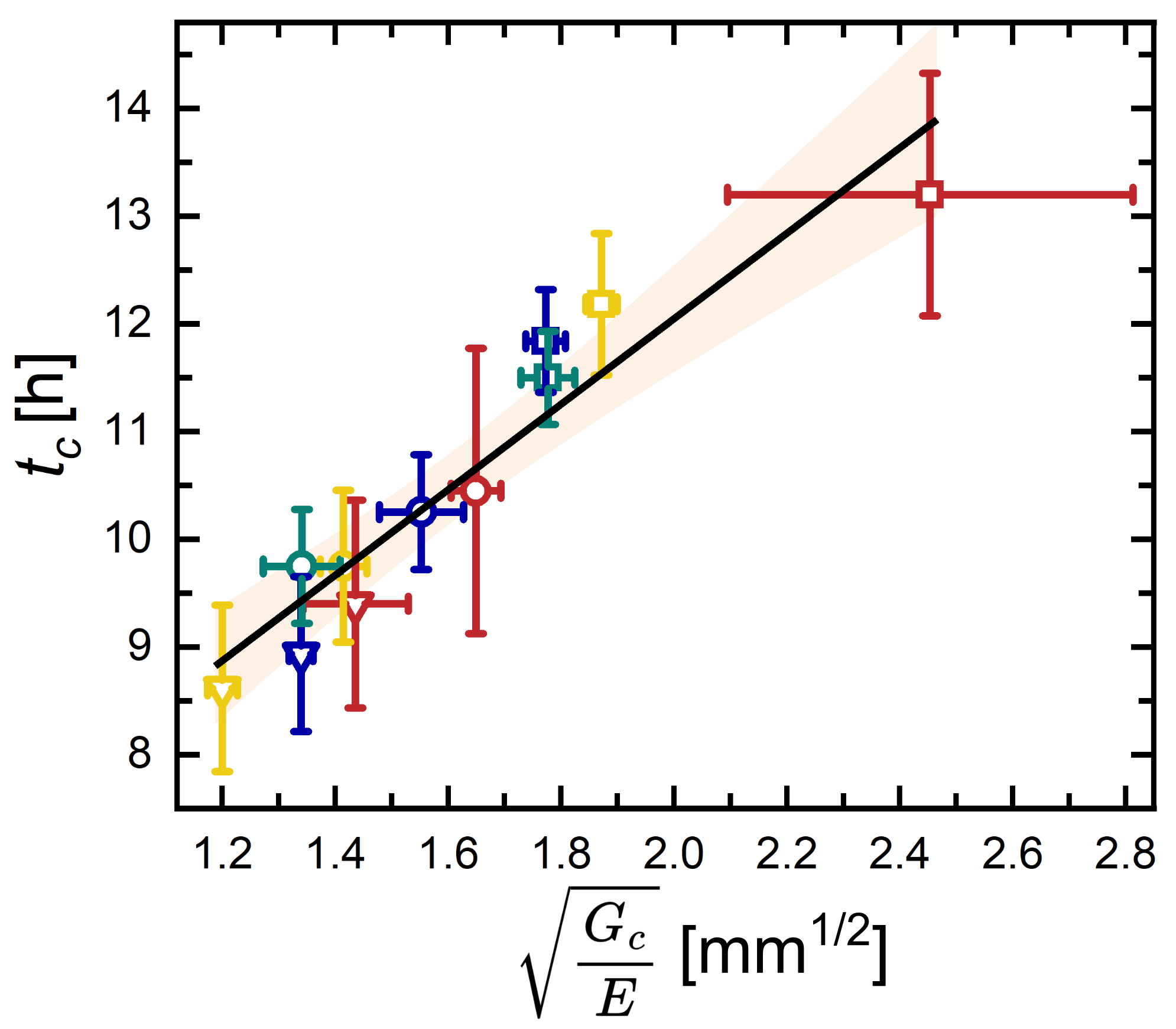}
\centering
    \caption{ Crack onset time $t_c$ \textit{vs.} $\sqrt{\frac{G_c}{E}}$. The solid line is a linear fit to the experimental data. Squares, circles and triangles respectively correspond to NaCl concentrations $C_A =$ 0, 2 and 4 mM. Red, blue, yellow and green symbols respectively correspond to clay concentrations $C_L =$ 2.0, 2.5, 3.0 and 3.5\% w/v.}
    \label{fig6}
\end{figure}
A pre-existing flaw is expected to nucleate a crack of length $a$ when the tensile stress reaches a critical value, $\sigma_c$. According to the Griffith's criterion for equilibrium crack propagation, the critical cracking stress at which the crack becomes active is given by~\cite{10.1002/9783527671922}:
\begin{equation}
    \sigma_c = \sqrt{\frac{G_c E}{\pi a}}
\end{equation}
where $G_c = 2\gamma + P$ is the fracture energy or critical energy release rate, $\gamma$ is the surface tension of the clay suspension, $P$ is the irreversible plastic work per unit area of crack propagation and $E$ is the sample elasticity at $t = t_c$.
For the crack to form at $t = t_c$, the tensile stress at the surface of the sample $<\sigma_{rr}(t=t_c)|_{r = R}>_z$ must be equal to the critical cracking stress $\sigma_c$. Therefore, using Eqns.~(3) and (4) and assuming $a \sim h$~\cite{10.1002/9783527671922}, we estimated the fracture energies, $G_c$, of our samples. Figure~\ref{fig5}(c) shows that $G_c$ decreases with increasing NaCl content, indicating decrease in sample ductility as sample elasticity increases. Even in the absence of evaporation, previous extension and oscillatory rheological experiments on Bentonite and Laponite clay suspensions have reported decreasing ductility as sample elasticity increases with time due to physical aging~\cite{10.1122/8.0000192,Hayes2022-te}. Our observation that a clay sample with accelerated aging is less ductile (is relatively more brittle) at $t = t_c$ and has a lower ability to undergo plastic deformation, is therefore in agreement with previous reports.% It is clear that the samples which cracked earlier are relatively less ductile at the time of crack onset.

The different terms in the expression for $<\sigma_{rr}(t)|_{r = R}>_z$ in Eqn.~(3) are plotted in supplementary material Fig.~S9. In agreement with previous data acquired from drying drops and layers of silica and latex particles~\cite{C7SM00985B,Giorgiutti-Dauphin2014,Chekchaki2013}, we see from supplementary material Fig.~S9 that the first term in Eqn.~(3) is the most dominant, therefore $<\sigma_{rr}|_{r=R, t = t_c}>_z$ $\approx
 \frac{V_EEt_c}{h}$. 
%The validity of this approximation has been previously demonstrated for other colloidal systems such as drying drops and layers of silica and latex particles~\cite{C7SM00985B,Giorgiutti-Dauphin2014,Chekchaki2013}. 
Retaining only the leading order term, we equated $<\sigma_{rr}|_{r=R, t = t_c}>_z$ with the critical cracking stress $\sigma_c$ in the Griffith's criterion (Eqn.~{4}) and estimate the crack onset time $t_c$ to be:
\begin{equation}
    t_c \approx \frac{h}{\sqrt{{\pi a}}V_E}\sqrt{\frac{G_c}{E}}
\end{equation}
 Indeed, our experimental data follows the correlation $t_c \propto \sqrt{\frac{G_c}{E}}$, as shown in Fig.~\ref{fig6}. The initiation and, consequently, the extent of cracking in consolidating clay suspension layers can therefore be controlled by tuning the rate of physical aging which would, in turn, determine the elasticity and ductility of the sample. We conclude that $\sqrt{\frac{G_C}{E}}$, obtained by combining the well-established linear poroelastic and Griffith fracture theories, is a key parameter governing the onset of cracks in drying clay suspensions and should also be applicable to drying drops and layers constituted by diverse isotropic colloidal gels. It follows from our study that the fracture energy of a sample can be estimated from desiccation experiments if the sample elasticity is known. Alternatively, the crack onset time may be predicted for a sample of predetermined initial thickness with well-characterized rigidity and ductility values.

 \section{\label{sec:4}Conclusions}
Aqueous clay suspensions exhibit a physical aging process wherein the suspension transforms from a viscoelastic liquid to a soft solid with elapsed time due to evolving {inter-particle }electrical double-layer interactions~\cite{doi:10.1021/acs.langmuir.8b01830,C0SM00590H,C2SM25731A}. We tuned the physical aging and, therefore, the rate of consolidation of Laponite clay suspensions by adding common salt or tetrasodium pyrophosphate (TSPP) to the aqueous medium. This allowed us to systematically investigate the combined effects of physical aging dynamics and solvent loss on the formation of desiccation cracks in the drying clay suspensions. For a fixed evaporation rate and adequate sample-boundary adhesion, we observed that clay samples consolidated faster in the presence of added common salt (NaCl) and were prone to earlier cracking. In contrast, addition of tetrasodium pyrophosphate (TSPP) suppressed the contribution of physical aging on the consolidation of the clay suspension and inhibited drying-induced cracks. The elasticity values of partially dried clay suspensions were next measured using microindentation in an atomic force microscope. By incorporating the Griffith's fracture criterion in the theory of linear poroelasticity, we obtained a correlation between the crack onset time, $t_c$, sample elasticity, $E$, and fracture energy, $G_c$, $t_c \propto \sqrt{\frac{G_c}{E}}$, which described our experimental data very well and allowed us to estimate the fracture energy of each sample. We concluded that adding NaCl to clay suspensions reduced their ductility and therefore their ability to deform and dissipate stresses. This led to early crack onset in these samples. 

Our study provides valuable guidelines to extract novel insights into the strength and load capacity of clay-based systems and is useful in formulating crack-resistant coatings and layers, for example, in the assembly of photonic bandgap crystals of colloids \textit{via} drying~\cite{10.1063/1.1737066}. Since Laponite clay suspensions form a nematic gel state when particle concentration is increased, it would be useful to derive a correlation between crack onset time and the mechanical properties of an anisotropic gel.  In practice, drying colloidal suspension layers of clay or paint would experience diurnal temperature and humidity variations. Controlled studies of desiccation crack formation under periodically varying temperature and relative humidity conditions would also be of great practical interest. Employing physics-informed machine learning algorithms~\cite{BOUKHTACHE2021106308} in the analysis of strain fields on the surface of a desiccating sample layer would result in accurate predictions of time and location of crack onset, thereby aiding in the prevention of catastrophic failure events.

\section{Supplementary Material}
 See the supplementary material for schematic illustrations of clay suspension microstructures, experimental chamber validation, crack onset time analyses, sample elasticity measurements, final crack patterns, experiments at varying evaporation rates, DIC analyses for different sample-boundary adhesion conditions, permeability and clay layer height calculations, and stress analyses.\\
 See supplementary video 1 for the temporal evolution of desiccation cracks in a clay suspension of concentration 3.0\% w/v and 4 mM NaCl. Supplementary video 2 shows the temporal changes in deformation fields as a function of elapsed time for different sample-boundary adhesion conditions. 

\section{\label{sec:x}Data availability statement}
The data that supports the findings of this study are tabulated in supplementary material table~S1. Additional raw data files are available from the corresponding author upon reasonable request.

\section{\label{sec:5}CRediT authorship contribution statement}
\textbf{VRS Parmar:} Investigation (lead),  Methodology (lead), Formal analysis (lead),  Writing– original draft (lead). \textbf{RB:}  Conceptualization (equal), Resources (lead),  Supervision (lead),  Writing– review \& editing (equal).

\section{\label{sec:6}Declaration of competing interest}
The authors have no conflicts to disclose.

\section{\label{sec:7}Acknowledgement}
The authors thank Yatheendran K. M. for help with AFM microindentation measurements, Shivprasad Patil for sharing his expertise in microindentation technique, Lucas Goehring for their expertise on desiccation cracks and Abhishek Ghadai for useful discussions related to digital image correlation experiments.

\bibliography{aipsamp}% Produces the bibliography via BibTeX.

\end{document}